\documentclass[twoside]{article}
\usepackage[T1]{fontenc}
\usepackage{graphicx}

\input{psfig.sty}
\def\newline{\hfil\break}

\def\bbbz{{\mathchoice {\hbox{$\sf\textstyle Z\kern-0.4em Z$}}
{\hbox{$\sf\textstyle Z\kern-0.4em Z$}}
{\hbox{$\sf\scriptstyle Z\kern-0.3em Z$}}
{\hbox{$\sf\scriptscriptstyle Z\kern-0.2em Z$}}}}  
\catcode`\@=11
\long\def\@makefntext#1{
\protect\noindent \hbox to 3.2pt {\hskip-.9pt
$^{{\eightrm\@thefnmark}}$\hfil}#1\hfill}		

\def\@makefnmark{\hbox to 0pt{$^{\@thefnmark}$\hss}}	

\def\ps@myheadings{\let\@mkboth\@gobbletwo
\def\@oddhead{\hbox{}
\rightmark\hfil\eightrm\thepage}
\def\@oddfoot{}\def\@evenhead{\eightrm\thepage\hfil
\leftmark\hbox{}}\def\@evenfoot{}
\def\sectionmark##1{}\def\subsectionmark##1{}}

\oddsidemargin=\evensidemargin
\newcounter{sectionc}\newcounter{subsectionc}\newcounter{subsubsectionc}
\renewcommand{\section}[1] {\vspace{12pt}\addtocounter{sectionc}{1}
\setcounter{subsectionc}{0}\setcounter{subsubsectionc}{0}\noindent
	{\tenbf\thesectionc. #1}\par\vspace{5pt}}
\renewcommand{\subsection}[1] {\vspace{12pt}\addtocounter{subsectionc}{1}
	\setcounter{subsubsectionc}{0}\noindent
	{\bf\thesectionc.\thesubsectionc. {\kern1pt \bfit #1}}\par\vspace{5pt}}
\renewcommand{\subsubsection}[1] {\vspace{12pt}\addtocounter{subsubsectionc}{1}
	\noindent{\tenrm\thesectionc.\thesubsectionc.\thesubsubsectionc.
	{\kern1pt \tenit #1}}\par\vspace{5pt}}
\newcommand{\nonumsection}[1] {\vspace{12pt}\noindent{\tenbf #1}
	\par\vspace{5pt}}

\newcounter{appendixc}
\newcounter{subappendixc}[appendixc]
\newcounter{subsubappendixc}[subappendixc]
\renewcommand{\thesubappendixc}{\Alph{appendixc}.\arabic{subappendixc}}
\renewcommand{\thesubsubappendixc}
	{\Alph{appendixc}.\arabic{subappendixc}.\arabic{subsubappendixc}}

\renewcommand{\appendix}[1] {\vspace{12pt}
        \refstepcounter{appendixc}
        \setcounter{figure}{0}
        \setcounter{table}{0}
        \setcounter{lemma}{0}
        \setcounter{theorem}{0}
        \setcounter{corollary}{0}
        \setcounter{definition}{0}
        \setcounter{equation}{0}
        \renewcommand{\thefigure}{\Alph{appendixc}.\arabic{figure}}
        \renewcommand{\thetable}{\Alph{appendixc}.\arabic{table}}
        \renewcommand{\theappendixc}{\Alph{appendixc}}
        \renewcommand{\thelemma}{\Alph{appendixc}.\arabic{lemma}}
        \renewcommand{\thetheorem}{\Alph{appendixc}.\arabic{theorem}}
        \renewcommand{\thedefinition}{\Alph{appendixc}.\arabic{definition}}
        \renewcommand{\thecorollary}{\Alph{appendixc}.\arabic{corollary}}
        \renewcommand{\theequation}{\Alph{appendixc}.\arabic{equation}}
        \noindent{\tenbf Appendix \theappendixc #1}\par\vspace{5pt}}
\newcommand{\subappendix}[1] {\vspace{12pt}
        \refstepcounter{subappendixc}
        \noindent{\bf Appendix \thesubappendixc. {\kern1pt \bfit #1}}
	\par\vspace{5pt}}
\newcommand{\subsubappendix}[1] {\vspace{12pt}
        \refstepcounter{subsubappendixc}
        \noindent{\rm Appendix \thesubsubappendixc. {\kern1pt \tenit #1}}
	\par\vspace{5pt}}

\newcommand{\textlineskip}{\baselineskip=13pt}
\newcommand{\smalllineskip}{\baselineskip=10pt}

\def\eightcirc{
\begin{picture}(0,0)
\put(4.4,1.8){\circle{6.5}}
\end{picture}}
\def\eightcopyright{\eightcirc\kern2.7pt\hbox{\eightrm c}}

\newcommand{\copyrightheading}[1]
	{\vspace*{-2.5cm}\smalllineskip{\flushleft
	{\footnotesize International Journal of Modern Physics C, #1}\\
	{\footnotesize $\eightcopyright$\, World Scientific Publishing
	 Company}\\
	 }}

\newcommand{\publisher}[2]{{\begin{center}\footnotesize\smalllineskip
	Received #1\\
	Revised #2
	\end{center}
	}}

\def\abstracts#1#2#3{{
	\centering{\begin{minipage}{4.5in}\baselineskip=10pt\footnotesize
	\parindent=0pt #1\par
	\parindent=15pt #2\par
	\parindent=15pt #3
	\end{minipage}}\par}}

\def\keywords#1{{
	\centering{\begin{minipage}{4.5in}\baselineskip=10pt\footnotesize
	{\footnotesize\it Keywords}\/: #1
	\end{minipage}}\par}}

\renewenvironment{thebibliography}[1]
        {\frenchspacing
	 \ninerm\baselineskip=11pt
         \begin{list}{\arabic{enumi}.}
        {\usecounter{enumi}\setlength{\parsep}{0pt}
	 \setlength{\leftmargin 12.7pt}{\rightmargin 0pt} 
         \setlength{\itemsep}{0pt} \settowidth
	{\labelwidth}{#1.}\sloppy}}{\end{list}}

\newcounter{itemlistc}
\newcounter{romanlistc}
\newcounter{alphlistc}
\newcounter{arabiclistc}

\newcommand{\fcaption}[1]{
        \refstepcounter{figure}
        \setbox\@tempboxa = \hbox{\footnotesize Fig.~\thefigure. #1}
        \ifdim \wd\@tempboxa > 5in
           {\begin{center}
        \parbox{5in}{\footnotesize\smalllineskip Fig.~\thefigure. #1}
            \end{center}}
        \else
             {\begin{center}
             {\footnotesize Fig.~\thefigure. #1}
              \end{center}}
        \fi}

\newcommand{\tcaption}[1]{
        \refstepcounter{table}
        \setbox\@tempboxa = \hbox{\footnotesize Table~\thetable. #1}
        \ifdim \wd\@tempboxa > 5in
           {\begin{center}
        \parbox{5in}{\footnotesize\smalllineskip Table~\thetable. #1}
            \end{center}}
        \else
             {\begin{center}
             {\footnotesize Table~\thetable. #1}
              \end{center}}
        \fi}

\def\@citex[#1]#2{\if@filesw\immediate\write\@auxout
	{\string\citation{#2}}\fi
\def\@citea{}\@cite{\@for\@citeb:=#2\do
	{\@citea\def\@citea{,}\@ifundefined
	{b@\@citeb}{{\bf ?}\@warning
	{Citation `\@citeb' on page \thepage \space undefined}}
	{\csname b@\@citeb\endcsname}}}{#1}}

\newif\if@cghi
\def\cite{\@cghitrue\@ifnextchar [{\@tempswatrue
	\@citex}{\@tempswafalse\@citex[]}}
\def\citelow{\@cghifalse\@ifnextchar [{\@tempswatrue
	\@citex}{\@tempswafalse\@citex[]}}
\def\@cite#1#2{{$\null^{#1}$\if@tempswa\typeout
	{IJCGA warning: optional citation argument
	ignored: `#2'} \fi}}

\def\pmb#1{\setbox0=\hbox{#1}
	\kern-.025em\copy0\kern-\wd0
	\kern.05em\copy0\kern-\wd0
	\kern-.025em\raise.0433em\box0}

\def\fnt#1#2{\footnotetext{\kern-.3em
	{$^{\mbox{\scriptsize #1}}$}{#2}}}

\def\fpage#1{\begingroup
\voffset=.3in
\thispagestyle{empty}\begin{table}[b]\centerline{\footnotesize #1}
	\end{table}\endgroup}

\def\runninghead#1#2{\pagestyle{myheadings}
\markboth{{\protect\footnotesize\it{\quad #1}}\hfill}
{\hfill{\protect\footnotesize\it{#2\quad}}}}
\font\tenrm=cmr10
\font\tenit=cmti10
\font\tenbf=cmbx10
\font\bfit=cmbxti10 at 10pt
\font\ninerm=cmr9

\font\eightrm=cmr8

\font\tit=cmbx10 scaled \magstep2   
 at 10truept 
\font\pic=cmr7  at 7truept  

\def\p1{\hbox to 0.5truecm{\hfill{\tit .}\hfill}}

\def\punt#1{\hbox to #1truecm{\leaders\hbox to 0.5truecm{{\tit .}\hfill}
\hfill}}
\def\cer#1{\hbox to #1truecm{\leaders\hbox to 0.5truecm{$\circ$\hfill}
\hfill}}
\def\p3{\hbox to 0.57735truecm{\hfill{\tit .}\hfill}}
\def\punlet#1{\hbox to 0.5truecm{{\tit .}{\pic #1}\hfill}}
\begin{document}
\runninghead{A. Bonelli \& S. Ruffo}{Modular transformations and random numbers}
\normalsize\textlineskip
\thispagestyle{empty}
\setcounter{page}{1}
\copyrightheading {Vol. 9, No. 4 (1998) 000--000}
\vspace*{0.88truein}
\fpage{1}

\centerline{\bf MODULAR TRANSFORMATIONS, ORDER-CHAOS TRANSITIONS}
\centerline{\bf AND PSEUDO-RANDOM NUMBER GENERATION}
\vspace*{0.37truein}
\centerline{\footnotesize ANTONIO BONELLI}
\vspace*{0.015truein}
\centerline{\footnotesize\it Dipartimento di Chimica}
\centerline{\footnotesize\it Universit\`a della Basilicata}
\centerline{\footnotesize\it via Nazario Sauro 85, I-85100, Potenza, Italy}
\centerline{\footnotesize\it E-mail: bonelli@ch1ris.cisit.unibas.it}
\vspace*{0.15truein}
\centerline{\footnotesize and}
\vspace*{0.15truein}
\centerline{\footnotesize STEFANO RUFFO}
\vspace*{0.015truein}
\centerline{\footnotesize\it Dipartimento di Energetica}
\centerline{\footnotesize\it Universit\`a di Firenze}
\centerline{\footnotesize\it via s. Marta 3, I-50139 Firenze, Italy}
\centerline{\footnotesize\it INFN, sez. Firenze}
\centerline{\footnotesize\it E-mail: ruffo@avanzi.de.unifi.it}

\vspace*{0.225truein}
\publisher{August 1, 1998}{August 7, 1998}
\vspace*{0.21truein}

\abstracts{Successive pairs of pseudo-random numbers generated by
standard linear congruential transformations display ordered patterns
of parallel lines. We study the ``ordered'' and ``chaotic'' distribution
of such pairs by solving the eigenvalue problem for two-dimensional 
modular transformations over integers. 
We conjecture that the optimal uniformity for pair distribution is obtained
when the slope of linear modular eigenspaces takes the value 
$n_{opt} = \mbox{maxint}(p /\sqrt{p-1})$, where $p$ is a prime number.
We then propose a new generator of pairs of independent pseudo-random
numbers, which realizes an optimal uniform distribution (in the ``statistical'' 
sense) of points on the unit square $(0,1] \times (0,1]$. The method can
be easily generalized to the generation of $k$-tuples of random numbers
(with $k>2$).  
}{}{}
\vspace*{10pt}
\vspace*{1pt}\textlineskip

\section{Introduction}
Consider the standard linear congruential pseudo-random number generator
(see Ref.\cite{gut} for a review on pseudo-random numbers generation)
\begin{equation}
x_{t+1}=gx_t \, \mbox{mod p},
\label{minimal}
\end{equation}
with $p=2^{31}-1$ and $g=7^5=16807$ (also called in this case
``minimal standard''~\cite{press}).
It is well known that successive pairs $(x_t,x_{t+1})$, normalized
in the interval $(0,1]$, show a pattern of parallel lines on the
square $S=(0,1]\times(0,1]$~\cite{Vattu}. This observation
is even more general, since this pattern is observed for all pairs
$(x_t,x_{t+l})$, with $l=1,\dots,p-1$\cite{DeM89}. However, the number
of parallel lines over which the pairs are distributed strongly depends
on the value of $l$. For instance, for $l=(p-1)/2$, the pairs are
distributed on a single line, while for $l=(p-1)/2 \pm 1$ the pairs
are distributed over so many parallel lines that the pairs could be
considered as randomly filling the square $S$. This
curious behavior has been called ``order'' (pairs distributed over a
few lines) to ``chaos'' (pairs distributed over several lines) transition
in Ref.~\cite{DeM89}. This remark is of course very important to
establish the properties of those correlations in pseudo-random number
generation which are due to the ``lattice'' structure created by
the properties of the integer field $Z_p=\{0,\dots,p-1\}$ over which
map (\ref{minimal}) is defined.

In this paper we analyse the pattern of parallel lines using the theory
of continuous bijective transformations defined by unimodular $2 \times 2$ 
matrices~\cite{Arn68} (also called in mathematical papers Continuous
Automorphisms of the Torus, CAT), which we recall in Section 2. 
We give a simple explanation of why the pairs are distributed 
either ``orderly'' or ``chaotically'', by studying the dependence on 
$l$ of the slope of Linear Modular Eigenspaces (LME), after solving the 
eigenvalue problem for CATs (see Sections 2 and 3).

By introducing appropriate probes of the uniformity of pair distributions
over the square $S$ (Section 4), we are able to guess the value of the
slope of the LME (and correspondingly the value of $l$) which gives
the most uniform pair distribution.

This study suggests the use of linear congruential pseudo-random number 
generators of the kind (\ref{minimal}) to generate pairs of independent
pseudo-random numbers, by appropriately choosing the initial pair
over the most uniform LME. We exploit this idea in Section 5, but we
discover that the requirement of optimal uniformity in the ``geometrical''
sense does not produce the best pair generator in the ``statistical''
sense. Hence, we modify the pair generator by introducing a variation
of the LME slope, which produces a better randomization of the successive
pairs. We compare the uniformity of the generated pairs with those
obtained with a commonly used pseudo-random number generator ({\it ran0}
in Ref.~\cite{press}).

In Section 6 we draw some conclusions.

\section{Modular transformations\label{sec:Intro}}
Continuous bijective transformations (isomorphisms) of the torus
(the square $S=(0,1] \times (0,1]$ with the opposite sides identified)
are maps defined by unimodular (unitary determinant) $2 \times 2$ 
matrices $A$ of integers~\cite{Arn68},
\begin{equation}
\left( \matrix{x(t+1)\cr y(t+1)\cr} \right)=
\left( \matrix{a & b\cr c & d\cr} \right)
\left( \matrix{x(t)\cr y(t)\cr} \right) \ \mbox{mod 1}~,
\label{auto}
\end{equation}
where $x,y$ are reals and $a,b,c,d$ integers and the $\mbox{mod 1}$
operation means taking the fractional part in $(0,1)$ of the real number.
In this paper we restrict to the study of periodic orbits, whose initial 
point, $(x(0),y(0))=(q/p,q'/p)$, is rational, with common denominator $p$. 
These orbits are then equivalently generated by the
transformation
\begin{equation}
\left (\matrix{x(t+1)\cr y(t+1)\cr}\right )=
\left(\matrix{a & b\cr c & d\cr}\right )
\left (\matrix{x(t)\cr y(t)\cr}\right ) \ \mbox{mod p}~,
\label{auto1}
\end{equation}
where $x,y$ are integers belonging to $[0,p-1]$ (the field $Z_p$), thanks 
to the $\mbox{mod p}$ operation.

Moreover, we restrict to  prime values of $p$, which allow for the 
existence of orbits with maximal period $p-1$. Let 
$\Delta = (\mbox{Tr} A)^2 -4$, if an integer $q= 0,\dots,p-1$ exists 
such that $q^2 = \Delta \, \mbox{mod p}$, then matrix $A$ admits 
two eigenvalues
\begin{equation}
\lambda_{\pm} = 2^{-1}(\mbox{Tr} A \pm q)~,
\end{equation}
where $2^{-1}$ is the multiplicative inverse of 2 in the field $Z_p$.
From these properties we can directly derive the equations for linear
modular eigenspaces (LME)
\begin{equation}
y = n_\pm x \, \mbox{mod p}~,
\label{line}
\end{equation}
where 
\begin{equation}
n_\pm = -b^{-1} (a - \lambda_\pm)~,
\label{npiu}
\end{equation}
and $(x,y) \in Z_p \times Z_p$.

A 'primitive root' of $p$, is a positive integer $g$ for which the smallest
value of $s$ such that $g^s = 1 \, \mbox{mod} \ p$ is $s=p-1$. It has been
observed~\cite{DeM89} that ordered pairs  $(g^n,g^{n+l})$ 
in $Z_p \times Z_p$ with varying $n=1,\dots,p-1$ and the fixed parameter
$l$ chosen in the set $\{1,\dots,p-1\}$ distribute ``orderly'' or ``chaotically''
on the lattice of points $Z_p \times Z_p$ depending on the chosen value of
$l$. This result is relevant for the understanding of correlations in
the generation of random numbers by modular transformations of integers.

These ordered pairs can also be seen as successive points over an orbit
of map (\ref{auto1}) with initial value $(g^k,g^{k+l})$ 
($k \in \{0,\dots,p-1\}$) belonging to the LME 
(\ref{line}). Therefore, choosing the value of $l$ and the primitive
root $g$ is equivalent to fixing the elements of matrix $A$ in (\ref{auto1})
and the initial condition, in such a way that the following consistency
relation is obeyed
\begin{equation}
bg^l-g+a=0~.
\end{equation}
In this paper we propose to study the properties of the pairs, or equivalently
of the orbits of (\ref{auto1}), as a function of the parameter $n_\pm$
which fixes the LME, and we establish a relation
of 'similarity' among these orbits after introducing discrete rotations
of the point lattice (Section 3). This allows us to overcome a 
difficulty encountered
in Ref.~\cite{DeM89}, where to the slightest variation of parameter $l$
corresponded a drastic change in the behavior of the orbit, from ``order''
to ``chaos'' and back, with no apparent regularity in the change. Using our
approach it is possible to classify the different behaviors of the orbits
according to the equation of the LME once $p$ and $g$ are given.

\section{Linear modular eigenspaces\label{sec:LME}}

The study of LME equations (see formula (\ref{line})) as the parameter 
$n_\pm$ is varied
is useful to determine the graph of linear periodic orbits of a specific
dynamical system fixed by the values of $p$ and $g$.
The value of $n_\pm$ varies in the set $\{0,1,\dots,p-1\}$. Eq.~(\ref{line})
can then be rewritten as
\begin{equation}
y = n_\pm x -h p~,
\label{line1}
\end{equation}
with $h=0,1,\dots,n_\pm-1$ and $[hp/n_\pm]+1 \leq x \leq [(h+1)p/n_\pm]$. 
This is the equation of an ensemble of $n_\pm$ parallel segments, whose
slope is $n_\pm$. If $n_\pm$ divides $p-1$, then the number of points
of the LME over each segment is the same.
Periodic orbits over a LME contain all points of the LME, but visit them
irregularly, jumping among the segments of the broken line (\ref{line1}).

Let us give an example. Consider the matrix
\begin{equation}
\left(\matrix{3 & 1\cr 2 & 1\cr}\right )~,
\label{ex1}
\end{equation}
whose eigenvalues are $\lambda_\pm = 2 \pm 2^{-1}q$, with $q^2=12$. Let
us consider the prime $p=13$ and the primitive root $g=6$. There are two 
values of $q$ whose square is $12$ in the field $Z_{13}$, i.e. $q=8$ and $q=5$;
we choose here $q=8$ and then $n_+ = 3$, $n_-=8$, $\lambda_+=6$ and
$\lambda_-=11$, the latter being also primitive roots. Taking the initial 
point on the LME, e.g.  $(x(0),y(0))=(1,3)$ one generates pairs which 
remain on LME, but jump among the 3 segments in which the LME
breaks up (see Fig.~1); the orbit has period 12. With an 
orbit of the same period one can obtain a more homogeneous filling
of the point lattice. 

\begin{figure}[ht]
\centerline{\psfig{figure=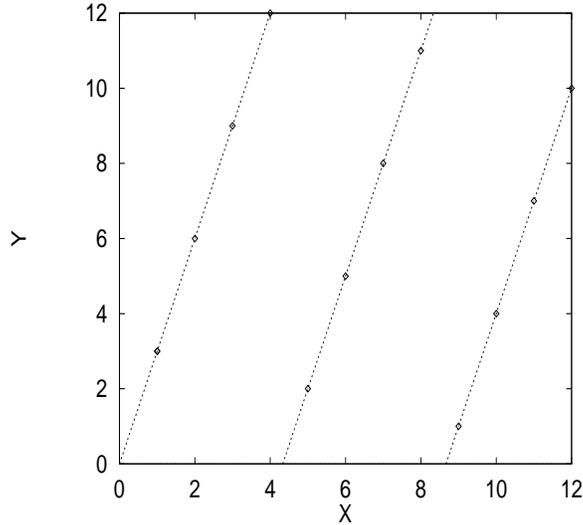,height=7cm,width=8cm}}
\caption{\label{n+=3} 
LME (dotted line) and a periodic orbit on it 
for $p=13$, $g=\lambda_+=6$, $n_+=3$.}
\end{figure}

The best one can obtain for these values of
$p$ and $g$ is for $n_+=5$. A possible choice of matrix
$A$ is obtained by the following procedure. First choose a value of $q$ which
respects the condition $q^2=(a+d)^2-4$; then, using equations
\begin{eqnarray}
2(a+n_{\pm}b) &=& a+d \pm q \\
ad -bc &=& 1~,
\end{eqnarray}
determine two of the elements of $A$ in terms of the others (e.g. the
diagonal terms $a$ and $d$). In our case one obtains the matrix
\begin{equation}
\left(\matrix{3 & 12\cr 11 & 1\cr}\right )~,
\label{ex2}
\end{equation}
whose eigenvalue $\lambda_+=11$ ($\lambda_-=6$) with
$q=5$ (the other choice of $q$ with respect to the previous example). 
In this case, since the prime $p=13=3^2+2^2$ is the sum of squares, the square 
$S_p=[0,p-1]\times[0,p-1]$ is tiled by $p$ squares of area $p$ (see Fig.~2), which realizes the most
uniform filling. 

\begin{figure}[ht]
\centerline{\psfig{figure=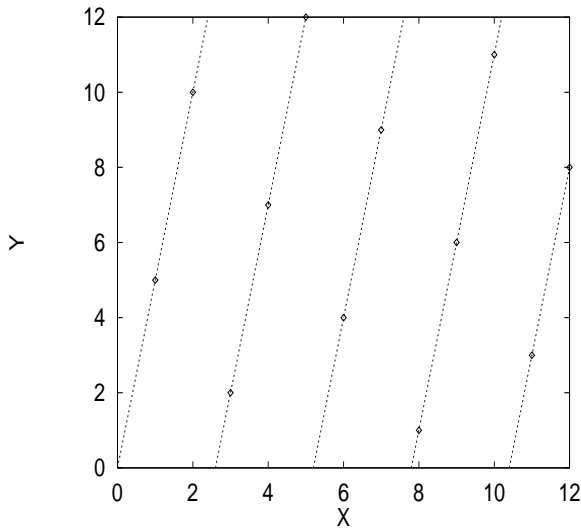,height=7cm,width=8cm}}
\caption{\label{n+=5} 
Same as Fig.~1 but $n_+=5$, which realizes a
more uniform filling of $S_{13}=[0,12] \times [0,12]$.}
\end{figure}

This recipe can be generalized to larger values
of $p$ which are sums of squares; the square of the diagonal of the 
tiling square is $2p$ and can be written in terms of $n_+$,
\begin{equation}
2p=n_+^2+1~,
\end{equation}
which gives $n_+=\sqrt{2p-1}$. This is the optimal choice for uniform
filling, but it is valid only for specific values of $p$. In the
following, we will find a more general rule valid for any $p$ (see formula
(\ref{best})).

Let us now introduce a 'similarity' relation among LMEs. We say
that two LMEs are similar if one can obtain a LME from the other by
rotating it around the point $(p/2,p/2)$ (intended as a real number) 
on the square $S_p$.
Let us consider the LMEs
\begin{equation}
y=m_i x \, \mbox{mod p}
\end{equation}
The similar LMEs are at most four, with $(m_0,m_1,m_2,m_3)$ such that
$m_0m_1=1 \mbox{mod p}$, $m_2=p-m_0$ and $m_3=p-m_1$. The LME with
slope $m_i$ is obtained rotating by $i \pi/2$ the
LME with slope $m_0$. The proof of this property is
trivial.    

\section{Uniformity probes:entropy and geometric index} 

In order to characterize the distribution of the points of the
periodic orbit on the square $S_p$, and its variation
with $n_\pm$, we define the 'entropy' of the periodic orbit as
follows.
Let us divide the square $S_p$ into square cells of side $\delta$.
Let $N_k$ be the number of points of the periodic orbit which
fall inside the $k-$th cell and $T_p=p-1$ the period. The entropy
is
\begin{equation}
\eta = - \sum_{k=1}^M w_k \ln w_k~,
\end{equation}
where $M=(p/\delta)^2$ is the total number of cells and $w_k = N_k/T_p$
is the fraction of points of the orbit inside the $k-$th cell.
To avoid the presence of empty cells (cells without points of the
orbit) in the homogeneous case, $\delta$ must be chosen such that 
$M=(p/\delta)^2=T_p=p-1$, then $\delta=p/\sqrt{p-1}$. As a consequence
\begin{equation}
0 < \eta \leq \ln (p-1)~.
\end{equation}
Another simple probe of homogeneity is the ratio of the sides, $L_{max}$
and $L_{min}$, of the elementary cell of the lattice of points formed by the
orbit in $S_p$,
\begin{equation}
r = \frac{L_{max}}{L_{min}}~.
\label{homo}
\end{equation}
Then $r \geq 1$ and the $r=1$ case corresponds to the most uniform
square elementary cell (which is reached only for the primes which
are sums of squares). The supremum value of $r$ is
$r_{sup}=p/\sqrt{2}$, which is attained when the points of the orbit
are all on a single segment (it is computed by considering the nearest
$S_p$ square).

In Fig.~3 we show the dependence of $\eta$ on $n$, the slope of
the LME. The chosen prime is $p=1999$ and the primitive root is $g=3$. 
The minimal entropy value, which is reached when the points 
of the orbit are placed on a single straight segment, is 
easily computed to be $\eta_{min}=1/2 \ln (p-1)$ (equal to $3.799\dots$
in this case) and is obtained when $n=1,1998$ ($n=1,p-1$ in general). 
The maximal entropy value $\eta_{max}=7.599\dots$ is reached 
for $n=45,844,1155,1954$; these values are related 
through the similarity relation introduced in Section 3.   

It is remarkable that the homogeneity ratio $r$ (see eq. \ref{homo})
gives a perfectly equivalent information, as shown in Fig.~4. 
We have verified that the agreement between the two homogeneity
probes gets better as the value of the prime $p$ increases.

\begin{figure}[ht]
\centerline{\psfig{figure=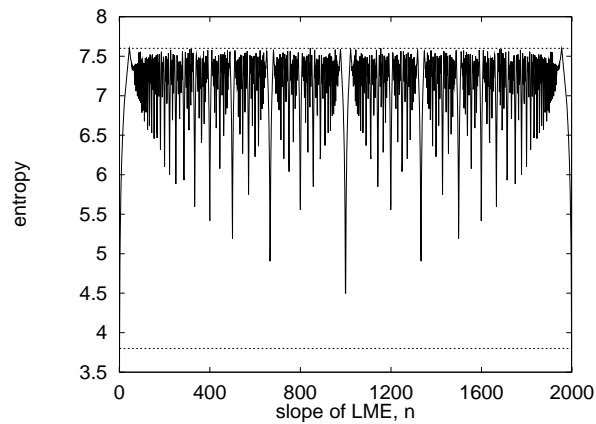,width=8cm}}
\caption{\label{entropy} 
Entropy of periodic orbits with $p=1999$ as a function of
$n$, also shown with dotted lines are the maximal, $7.599\dots$,
and minimal $3.799\dots$ entropies.}
\end{figure}

\begin{figure}[ht]
\centerline{\psfig{figure=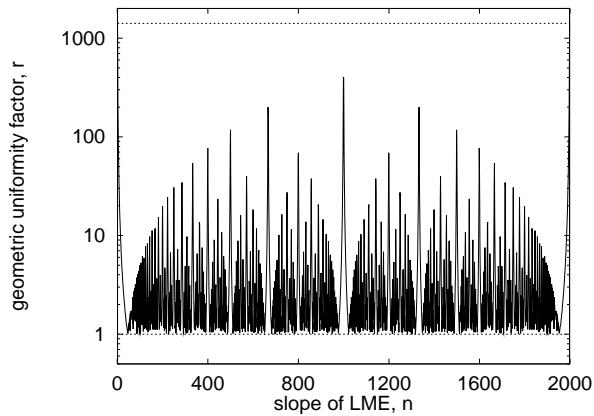,width=8cm}}
\caption{\label{geometric} 
Geometric uniformity factor $r$ for $p=1999$ 
as a function of $n$. The supremum $r_{sup}= 1413.506 \dots$ is 
reached for $n=1$ and for $n=1998$ when the orbit collapses onto
a segment. The supremum and the minimal $r=1$ values are shown by
the dotted lines.}
\end{figure}

Which is the optimal $n_{opt}$ value corresponding to the highest
homogeneity? A simple heuristic argument suggests that it must
be of the order of the side of the elementary square 
$\delta = p/\sqrt{p-1}$. In fact, numerical experiments suggest,
with high degree of confidence, the following conjecture
\begin{equation}
n_{opt} = \mbox{maxint} (\frac{p}{\sqrt{p-1}}).
\label{best}
\end{equation}
Fig.~5 shows the agreement of this conjectured value with the
one determined numerically. Of course, the similarity conditions
relate this value with three others.

\begin{figure}[ht]
\centerline{\psfig{figure=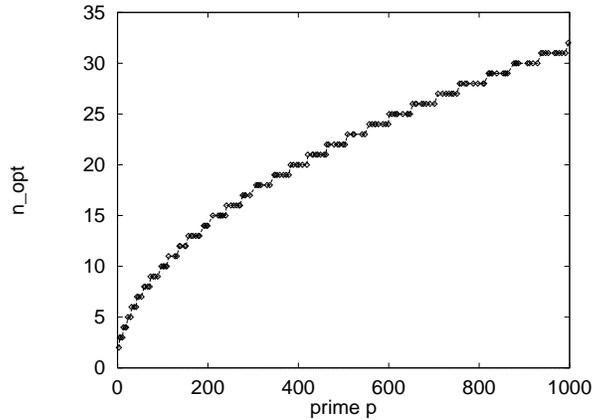,width=8cm}}
\caption{\label{nopt} 
Conjectured (line) and numerically determined (points) $n_{opt}$
as a function of the prime $p$.}
\end{figure}

\section{Random number generation}
One can use the transformation defined in (\ref{line}) by
the LME to generate recursively points at random with coordinates
$(X,Y)$ in the square $S=(0,1] \times (0,1]$ as follows
\begin{eqnarray}
X(t) &=& \frac{X(0)g^{t}}{p} \; \mbox{mod 1} \nonumber\\
Y(t) &=& \frac{g^l X(t)}{p} \; \mbox{mod 1}~,
\label{map}
\end{eqnarray}
once one has chosen the primitive root $g$ and the value $l$
such that $g^l=n_{\pm}$. $X(0) \in Z_p$ is the seed of the generator and
can always be written in terms of the primitive root $X(0)=g^{t_0}$
with $t_0 \in Z_p$.

In the previous section we have found the value of $n_{\pm}$ 
(\ref{best}), and consequently of $l$, which, for fixed $g$, 
gives the most uniform distribution of points on the square 
$S$ if the full orbit with initial value $(X(0)=g^{t_0},Y(0))$ 
is generated. 
We want now to test whether the recursion (\ref{map}) can also be used
as an efficient generator of random pairs $(X(t),Y(t))$ with
the appropriate choices of $g$ and $l$.
To test the efficiency we have used a method recently developed
in Ref.~\cite{robnik} to characterize ergodicity of tranformations
of the plane into itself. These authors introduce a {\it random model}
where the square $S$ is divided into $N$ cells and then points are
thrown into the cells each at a time with no correlations. 
At each step $t$ of the process one has equal probability 
$\pi=N^{-1}$, of choosing any cell. One can then define the probability
distribution $P_t(k)$ that $k$ cells are occupied at step $t$. The first
two moments of $P_t(k)$ are computed in~\cite{robnik}
\begin{eqnarray}
\rho(t) &=& \sum_{k=1}^N (k \pi) P_t(k) = 1 - (1 - \pi)^t \\
S(t) &=& \sum_{k=1}^N (k \pi)^2 P_t(k) = 1-(2-\pi)(1-\pi)^t+
(1-\pi)(1-2\pi)^t~.
\label{mom}
\end{eqnarray}
The first moment $\rho(t)$ is simply the density of occupied cells
at step $n$ and can be approximated for large $N$ as
\begin{equation}
\rho(t) \approx 1 - \exp \left( -\frac{t}{N} \right).
\end{equation}
The second moment gives the standard deviation $\sigma (t) = 
\sqrt{S(t) - \rho^2(t)}$ at step $t$.

In Fig. 6 we plot the numerically computed deviation with respect
to the theoretical result of the {\it random model} (\ref{mom})
for the random number generator (\ref{map}) with $p=2^{31}-1$,
$g=7^5$ and the optimal value $n_{opt}$ (\ref{best}) for the slope,
together with the deviation for a commonly used random number
generator, {\it ran0}~\cite{press}. The $\pm \sigma(t)$ values are
also reported and show that generator (\ref{map}) systematically
overestimates the theoretical curve, while {\it ran0} fluctuates
statistically around the theoretical value remaining within the
$\pm \sigma(t)$ region.

\begin{figure}[ht]
\centerline{\psfig{figure=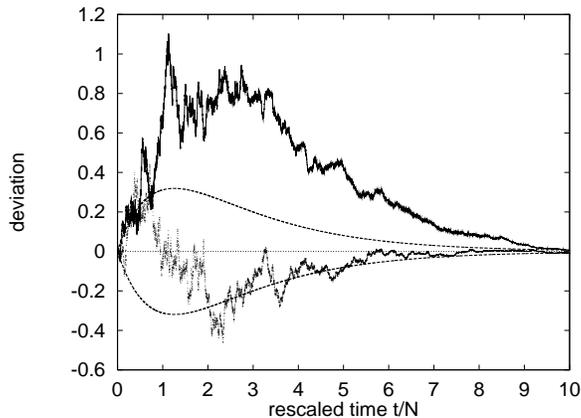,width=8cm}}
\caption{\label{comp1} 
Deviations from the theoretical curve of the density of occupied
cells in the {\it random model} for random series generated with
map (\protect\ref{map}) (full line) and {\it ran0} (dotted line). The 
$\pm \sigma(t)\cdot 10^3$ lines are also reported (broken lines).
$N=10^6$.}
\end{figure}

Our generator (\ref{map}) has therefore the tendency to produce
points on the square $S$ which overfill different cells with respect
to the random filling. This must be an effect of the temporal 
correlations which are present in (\ref{map}) and tend to distribute
points on parallel straight lines.
Therefore, we have modified (\ref{map}) imposing a variation at each
step of the slope of these lines. Hence, the generator is changed 
as follows
\begin{eqnarray}
X(t) &=& \frac{X(0)g^{t}}{p} \; \mbox{mod 1} \nonumber\\
r(t) &=& \frac{r(0)g^{t}}{p} \; \mbox{mod 1} \label{map1}\\
Y(t) &=& \frac{r(t) X(t)}{p} \; \mbox{mod 1}~ \nonumber,
\end{eqnarray}
with the seed $(X(0),r(0)) \in Z_p \times Z_p$. As previously, the 
seed can always be written in terms of the primitive root
$(X(0)=g^{t_0},r(0)=g^{s_0})$ with $t_0$ and $s_0$ belonging to $Z_p$.

The randomization of the slope $r$ reduces the correlations and now the
generated random series falls within the error curves 
$\pm \sigma(t)$ (see Fig.~7), showing the same quality of
{\it ran0}.

\begin{figure}[ht]
\centerline{\psfig{figure=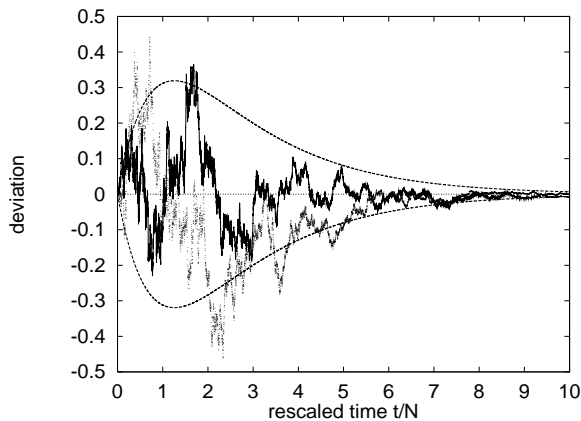,width=8cm}}
\caption{\label{comp2} 
Same as Fig.~6 where (\protect\ref{map}) has been substituted by 
(\protect\ref{map1}).}
\end{figure}

Although we do not claim here that generator (\ref{map1}) might be
competitive with others, more widely used, random number generators,
we are convinced to have revealed an important feature of
modular generators, i.e. cell overfilling tendency.
The recipe we have proposed to balance this effect is not be unique
and other solutions are certainly possible.

\section{Conclusions}
We have studied the geometrical properties of the orbits of modular 
transformations on integers. We have solved a puzzle proposed in
Ref.~\cite{DeM89} related to an ``order'' to ``chaos'' transition as
a parameter of the transformation is varied by showing that indeed the true
control parameter is the slope of linear modular eigenspaces; we have
then determined which of the values of this latter control parameter
give an ``ordered'' (i.e. distributed over a few segments) or 
``chaotic'' (i.e. spread over the lattice) orbit by looking at uniformity
probes.

Particular attention has been devoted to uniform (``chaotic'') 
orbits, which we have characterized using entropic and geometric
parameters. Moreover, we have verified that, during the time evolution,
the most ``geometrically'' uniform orbits do not correspond in general 
to the most ``statistically'' uniform, as defined through a
specific model introduced in Ref.~\cite{robnik}. 

With the aim of generating statistically uniform orbits we have introduced a
new generator of pseudo-random pairs, where the slope of the linear modular
eigenspace is varied during time evolution. 
The recipe we propose to generate pairs $(X,Y)$ of pseudo-random real 
(32-bit floating point) numbers with uniform distribution in the 
square $S=(0,1]\times(0,1]$ is by following recursion ($t=0,1,\dots$)
\begin{eqnarray}
X(t) &=& \frac{X(0)g^{t}}{p} \; \mbox{mod 1} \nonumber\\
r(t) &=& \frac{r(0)g^{t}}{p} \; \mbox{mod 1} \\
Y(t) &=& \frac{r(t) X(t)}{p} \; \mbox{mod 1}~ \nonumber,
\end{eqnarray}
where $p$ is a prime (e.g. the typical value $p=2^{31}-1$), $g$ a primitive
root (e.g. $g=7^5=16807$) and the seeds $X(0),r(0)$ are integers from 
1 to $p-1$.
The comparison of this generator with others commonly used is encouraging, 
as far as the uniformity properties are concerned. The method can also be 
easily extended to generate $k$-tuples of random numbers in $k$-dimensional 
spaces.

\nonumsection{Acknowledgements}

We thank Mario Rasetti for having suggested this topic of research and
for many useful discussions. Part of this work was performed while the
authors were visiting the Institute for Scientific Interchange in Torino,
which we thank for financial support. We also thank the anonymous referee
for having strongly contributed to improve (we hope) the readability
of this paper. 

\nonumsection{References}


\vspace*{-0.25cm}

\begin{thebibliography}{99}

\bibitem{gut}
F. Gutbrod, {\it New trends in pseudo-random number generation},
Annual Reviews of Computational Physics VI (World Scientific), to be
published.

\bibitem{press}
W.H. Press, S. Teukolsky, W.T. Vetterling and B.P. Flannery, {\it
Numerical Recipes}, Cambridge (1992).

\bibitem{Vattu}
I. Vattulainen, K. Kankaala, J. Saarinen and T. Ala-Nissila,
{\it Comp. Phys. Commun.}, {\bf 86}, 209 (1995).

\bibitem{DeM89}
De Matteis A., {\it Order out of chaos in arithmetic}, in G. Maino, L.
Fronzoni and M. Pettini eds., Dynamical simmetries and chaotic behaviour
in physical systems, World Scientific, (1989) and Refs. therein.
 
\bibitem{Arn68}  
Arnold V.I. and Avez A., {\it Ergodic problems of classical mechanics}, 
Benjamin, New York (1968)


\bibitem{robnik}
M. Robnik, J. Dobnikar, A. Rapisarda, T. Prosen and M. Petkovsek,
{\it J. Phys. A}, {\bf 30}, L803 (1997).



\end{thebibliography}
\end{document}